# Why Are We Obsessed with "Understanding" Quantum Mechanics?

Stephen Boughn*
Departments of Physics and Astronomy, Haverford College, Haverford PA

**Preface**

Richard Feynman famously declared, "I think I can safely say that nobody really understands quantum mechanics."[1] Sean Carroll lamented the persistence of this sentiment in a recent opinion piece entitled "Even Physicists Don't Understand Quantum Mechanics: Worse, they don't seem to want to understand it."[2] Quantum mechanics is arguably the greatest achievement of modern science and by and large we absolutely understand quantum theory. Rather, the "understanding" to which these two statements evidently refer concerns the ontological status of theoretical constructs.[3] For example, "Do quantum wave functions accurately depict physical reality?" The *quantum measurement problem* represents a collection of such queries and the conundrums to which they lead. Carroll offers two unsolved quantum riddles: 1) How is it that a quantum object exists as a superposition of different possibilities when we're not looking at it and then snaps into just a single location when we do look at it? and 2) Is the quantum mechanical wave function a complete and comprehensive representation of the world or does it have no direct connection with reality at all? According to Carroll,

> Until physicists definitively answer these questions, they can't really
> be said to understand quantum mechanics thus Feynman's lament.
> Which is bad, because quantum mechanics is the most fundamental
> theory we have, sitting squarely at the center of every serious attempt
> to formulate deep laws of nature. If nobody understands quantum
> mechanics, nobody understands the universe.

---

* sboughn@haverford.edu

[1] The 1964 Messenger Lectures at Cornell University, Lecture #6.
[2] NY Times, Sunday Review, Sept. 7, 2019
[3] Unlike Carroll, l doubt that Feynman was bothered by this lack of "understanding". See Section 3, *Final Remarks.*

On the other hand, these questions have been pondered since the dawn of quantum mechanics nearly a century ago and if they haven't been definitively answered by now, perhaps the source of the quandary is not with the answers but with the questions themselves. Inarguably, during that time numerous physicists and even more philosophers endeavored to "understand" quantum mechanics in this way. Some have claimed success; however, I doubt that many would agree. Most physicists are content with foregoing such metaphysical issues, falling back on Bohr's pragmatic Copenhagen interpretation, and then get on with the business of doing physics.

I confess that during my student days, and even thereafter, I was mightily bothered by these quantum mysteries and enjoyed spending time and effort worrying about them. Of course, as Carroll also laments, I avoided letting this avocation interfere with my regular physics research, otherwise, my academic career undoubtedly would have suffered.[4] As I approached the twilight of my career (I'm now retired), I happily resumed my ambition to "understand quantum mechanics" and have ended up writing several papers on the subject.[5] Furthermore, as others before me, I now proudly profess that I finally understand quantum mechanics ☺. Even so, I'm somewhat chagrinned that my understanding is essentially the same as that expressed by Niels Bohr in the 1930s, minus some of Bohr's more philosophical trappings.[6] Some have criticized my epiphany with remarks to the effect that I am too dismissive of the wonderful mysteries of quantum mechanics by relegating its role to that of an algorithm for making predictions while at the same time too reverent of insights provided by classical mechanics. I've come to believe that, quite to the contrary, those who still pursue an understanding of Carroll's quantum riddles are burdened with a classical view of reality and fail to truly embrace the fundamental quantum aspects of nature. In the remainder of this essay I'll illustrate this point of view with four famous quantum conundrums and leave it to the reader to conjure more examples.

---

[4] I'm reminded of the adage on the wall of the pool room in my undergraduate dining hall. "To play a good game of billiards is the mark of a gentleman. To play too good a game of billiards is the mark of an ill-spent youth."

[5] I'll spare you a listing and will reference only one, my overall view of the philosophical foundations of physics. (Boughn 2019)

[6] Heisenberg once remarked, in all due reverence, that Bohr was "primarily a philosopher, not a physicist" (Pais 1991).

1. **The Copenhagen Interpretation**

I'm sure that most of you are familiar with the "Copenhagen interpretation" of quantum mechanics and so I will not elaborate on its details. However, I need to address some of the many misstatements attributed to it.[7] My understanding of Bohr's interpretation comes primarily from a 1972 paper by Henry P. Stapp[8] and the reader is referred to that paper as well as another by Don Howard (2007) for an account of the mythology of the Copenhagen interpretation. Contrary to the claims of some, Bohr did not assert that quantum mechanics provides a complete description of an individual system but rather that (in Stapp's words) "…quantum theory provides a complete account of atomic phenomena…", that is, "…no theoretical construction can yield experimentally verifiable predictions about atomic phenomena that cannot be extracted from a quantum theoretical description." Neither did Bohr subscribe to the notion of the collapse of the wave function (that came from von Neumann and Dirac) nor did he believe in a privileged role for the observer. Bohr did not maintain that classical mechanics was a necessary component of the foundations of quantum mechanics, on the contrary, he claimed that all experimental apparatus are, in principle, subject to the laws of quantum mechanics. The quantum-classical divide to which he referred was simply the realization that the description of experiments must be in common language that can be comprehended by scientists and technicians alike. A point he made time and time again: it is the quantum of action that necessitates the probabilistic character of quantum mechanics. In fact, the statistical nature of quantum mechanics and the resulting conflicts with a classical notion of reality is in large part responsible for the quantum measurement problem and Carroll's quantum riddles. I'll refrain from commenting further on Bohr's interpretation except to the extent necessary in the following discussions of four quantum conundrums.

---

[7] "The Copenhagen Interpretation" was not a label approved by Bohr. Heisenberg introduced it in 1955.

[8] In fact this paper (Stapp 1972), which I read as a graduate student, has had a profound effect on my general view of the philosophical foundations of physics. (Boughn 2019)

## 2. Four Quantum Conundrums

The following four related questions are associated with the quantum measurement problem and all are of the sort for which Carroll insists we have definitive answers if we are ever "to formulate deep laws of nature". The first is the famous Einstein-Podolsky-Rosen "paradox", which many consider resolved; however, it presents a clear-cut example of how it is that the burden of a classical view of reality can seemingly confound the understanding of quantum mechanics.

*2.1 The Einstein-Podolsky-Rosen Paradox*

The 1935 paper by Einstein, Podolsky, and Rosen (EPR) expressed those authors dissatisfaction with the then current formulation of quantum mechanics. The conclusion was neither that quantum mechanics yields incorrect predictions nor that quantum mechanics is nonlocal, i.e., violates special relativity. Rather it was that quantum mechanics "…does not provide a complete description of the physical reality." The EPR manuscript was written in Einstein's absence by Podolsky and Einstein was not happy with it. He chose to couch his objection on the apparent action at a distance implied by quantum mechanics as will be discussed in the next section. Nevertheless, Einstein did agree with the conclusion that quantum mechanics does not provide a complete description of a single system.[9] The argument in EPR was a rather byzantine analysis involving a definition of elements of reality; however, in the end the conclusion was that both the position and the momentum of a particle were elements of *physics reality*. The hidden assumption running through the paper is that a single quantum system, in this case a single particle, has a well-defined location and momentum at each instant of time whether or not these quantities are observed. This is the essence of a classical world view. Without that assumption, the argument collapses.

A year later, Einstein (1936) offered a resolution to the EPR paradox, the ensemble interpretation of quantum mechanics. As Einstein described it:

> The Ψ function does not in any way describe a condition which
> could be that of a single system; it relates rather to many systems,
> to 'an ensemble of systems' in the sense of statistical mechanics…

---

[9] I agree that quantum mechanics does not provide a complete description of the physical world but for an entirely different reason. (See Section 2.4 below.)

> Such an interpretation eliminates also the [EPR] paradox recently demonstrated by myself and two collaborators…

The ensemble interpretation, while minimalist, is considered by some to be the most reasonable interpretation of quantum mechanics (e.g., see Ballentine 1970). While Einstein admitted that "such an interpretation eliminates also the [EPR] paradox", he added "To believe this is logically possible without contradiction; but, it is so very contrary to my scientific instinct that I cannot forego the search for a more complete conception". In other words, he would still strongly suspect that quantum mechanics is incomplete. The ensemble interpretation seems to be immune to the EPR conclusion that quantum mechanics cannot provide a complete description of a single system by requiring that quantum mechanics only apply to ensembles of systems. However, the specter of classical reality is still present for what does the ensemble represent if not a collection of single particles each with a definite position and momentum. Ballentine's presentation of the ensemble interpretation reinforces this notion. If so, it's hard to imagine how Einstein or Ballentine could accept quantum mechanics as a complete description of reality and thereby resolve the EPR paradox.[10]

*2.2 Bell's Theorem, Entanglement, and Quantum Non-locality*

Many credit the EPR paper with bringing the notion of entangled quantum states and action at a distance to the attention of the physics community; however, the major figures in physics at the time were already arguing about entanglement and Einstein had been bothered by action at a distance since the 1927 Solvay conference (Howard 2007). As I mentioned above, Einstein was not happy with the EPR paper. In a letter to Schrödinger written a month after the appearance of EPR he wrote (Howard 2007), "…the little essay… has not come out as well as I really wanted; on the contrary, the main point was, so to speak, buried by the erudition." In that same letter Einstein chose to base his argument for incompleteness on what he termed the "separation principle" according to which the *real state of affairs* in one part of space cannot be affected instantaneously or superluminally by events in a distant part of space. Suppose *AB* is the joint state of two systems, *A* and *B*, that interact and subsequently move away from each other to different locations. (Schrödinger (1935) coined the term *entangled* to describe such a joint state.) In his letter to Schrödinger, Einstein explained (Howard 2007)

> After the collision, the real state of (*AB*) consists precisely of the real state *A* and the real state of *B*, which two states have nothing to do with one another. The real state of *B* thus cannot depend upon the kind of measurement I carry out on *A* [separation principle]. But then for the same state of *B* there are two (in general arbitrarily many) equally justified [wave functions] $\Psi_B$, which contradicts the hypothesis of a one-to-one or complete description of the real states.

---

[10] As I pointed out in the *Introduction*, Bohr did not assert that quantum mechanics provides a complete description of an individual system; however, why couldn't one apply a statistical interpretation to a single system. For example, I would be happy to place a bet on the outcome of a quantum transition in a single system if the odds were heavily in my favor (using a Bayesian notion of probability).

His conclusion was again that quantum mechanics cannot provide a complete description of reality. In this argument Einstein moves from the classical notion that the position and momentum of a particle are elements of physics reality to the presumption that the quantum wave function of a particle must be unique if it is to provide a complete description of the *real physical state* of the particle. Quantum mechanics does not satisfy this requirement thus it cannot provide a complete description of the system. This also involves a classical view of reality; it simply replaces the words "the position and momentum of a particle are elements of physical reality" with "quantum wave functions are elements of physical reality". Quantum mechanics doesn't allow this characterization.

Einstein's argument invokes action at a distance, "the real state of *B* thus cannot depend upon the kind of measurement I carry out on *A*", which had bothered him since 1927. This brings us to John Bell's 1964 paper, "On the Einstein Podolsky Rosen Paradox". Bell was interested in "hidden variable" theories, classical theories with unknown parameters that could faithfully reproduce the predictions of quantum mechanics. His conclusion was that the only way this could be accomplished was with superluminal interactions, the "spooky action at a distance" that vexed Einstein. Bell chose a spin singlet system as an example of an entangled state. In his notation, let $A(\hat{a}, \lambda)$ and $B(\hat{b}, \lambda)$ represent the results of the spin measurements of particles 1 and 2 in the $\hat{a}$ and $\hat{b}$ directions respectively. Bell's condition of *locality* is that the correlation of any two measurements, $P(\hat{a}, \hat{b})$, be given by $P(\hat{a}, \hat{b}) = \int d\lambda\, \rho(\lambda) A(\hat{a}, \lambda) B(\hat{b}, \lambda)$ where $\lambda$ represents the hidden variables with statistical distribution $\rho(\lambda)$. A moment's thought should convince one that the product of measurement outcomes in this expression precludes the types of correlations (quantum interference) that quantum mechanics demands and so it's not surprising that the subsequent analysis of a sequence of such measurements is in conflict with the predictions of standard quantum mechanics. The expression for $P(\hat{a}, \hat{b})$ is clearly motivated by a classical view of reality, but to be fair the purpose of Bell's paper was to investigate the possibility of a classical, hidden variable theory. His conclusion was that any such theory could not be Lorentz invariant, i.e., it must be non-local. However, in a subsequent paper (Bell 1975), he formalized this "notion of local causality" and, from the violation of a set of derived inequalities, concluded that not only is quantum mechanics nonlocal[11] but also there exits a "gross nonlocality of nature". Again, here is a quantum riddle that rests wholly on the persistence of a classical view of reality.

*2.3 Wave Function Collapse and the Paradox of Schrödinger's Cat*

Recall Carroll's first quantum riddle, *How is it that a quantum object exists as a superposition of different possibilities when we're not looking at it and then snaps into just a single location when we do look at it?* This snapping into a single location is an example of what's referred to as collapse of the wave function and Carroll wants to know *why* this happens. There have been attempts to modify quantum mechanics so that such a collapse is a theoretical prediction; however, there are no experimental observations that suggest such a modification is necessary. Standard quantum mechanics is self-consistent and its predictions are in agreement with all current observations. Then why even ask the question? As far as I can tell, the only reason is a desire to attach a classical reality to the quantum wave function as did Einstein in his argument for the incompleteness of quantum mechanics. If one forgoes this notion, there is absolutely no reason to pursue the matter further.

A related issue is the paradox of Schrödinger's cat (Schrödinger 1935). In Schrödinger's 1935 gedanken experiment, a cat is penned up in a chamber with a radioactive substance that is monitored by a Geiger counter. The half-life of the substance is such that after one hour there is a 50% chance that a single radioactive atom will have decayed in which case the counter discharges. If so, then a relay releases a hammer that shatters a flask of hydrocyanic acid and the cat dies. If the counter does not discharge, the flask is not broken and the cat lives. After one hour, the wave function of the entire system expresses this situation by having equal parts of an alive cat/undecayed atom and a dead cat/decayed atom and thus the "indeterminacy originally restricted to the atomic domain becomes transformed into macroscopic indeterminacy" (Schrödinger 1935). When the box is opened and the cat plus atom system observed, the wave function collapses into one of these two states. This is absurd. Surely, the cat is either alive or dead before the box is opened even if we don't know which is the case. A way out of this conundrum is to claim that the wave function collapses as soon as the (macroscopic) Geiger counter clicks. So perhaps the collapse mechanism does make sense after all. On the other hand, the predictions of standard quantum mechanics are completely consistent with what we observe and until we have experimental evidence to the contrary, why purse the issue further. The only reason I can think of is to satisfy our notion of classical reality.

---

[11] I've always found it curious that the fact that standard quantum mechanics forbids superluminal signaling hasn't disabused those who assert quantum nonlocality of their claims.

*2.4 The Quantum-Classical Divide*

As a final example, let's consider the question of what constitutes an observation that supposedly causes the superposition of positions of a particle to "suddenly snap into just a single location." As Carroll puts it, "Why are observations special? What counts as an 'observation', anyway?" According to the Copenhagen interpretation, or so the standard argument goes, the act of measurement or observation must be described in terms of classical physics. Because classical physics is not (generally) couched in probabilistic terms, the classical measurement of a quantum system yields the specific result required by the probabilistic interpretation of the wave function. While possibly resolving the measurement dilemma, this explanation immediately raises two related questions: 1) When and where does the classical measurement occur, i.e., where is the *quantum-classical divide*? and 2) How does one describe the physical interaction across this divide? In posing these two questions, the specter of the notion of classical reality has again made its unwelcome appearance. The conviction that every interaction involves some physical process that is located at a specific time and place unquestionably arises from a classical view of reality.

A related conundrum is why is it necessary to revert to classical physics in order to understand quantum mechanics. Why didn't Bohr seem to worry about the unseemly merger of classical and quantum formalisms? I suspect the answer is because he didn't actually consider that any such merger is required. While Bohr endeavored to be extremely careful in expressing his ideas, his prose is often obscure. However, consider his following brief description of a measurement (Bohr 1963, p. 3):

> The decisive point is to recognize that the description of the experimental arrangement and the recordings of observations must be given in plain language, suitably refined by the usual terminology. This is a simple logical demand, since by the word 'experiment' we can only mean a procedure regarding which we are able to communicate to others what we have done and what we have learnt.

Nowhere in this description does he refer to classical physics. To be sure, Bohr often used the term "classical" in describing measurements but as Camilleri and Schlosshauer (2015) point out, "Bohr's doctrine of classical concepts is not primarily an interpretation of quantum mechanics…but rather is an attempt by Bohr to elaborate an epistemology of experiment." That the prescription for how to perform an experiment must be subsumed within the theoretical formalism of quantum mechanics follows only by analogy with classical physics. Again, it seems that a classical worldview lurks behind this quantum conundrum.[12] While I just characterized the requirement that experimental prescriptions be included in the theoretical formalism as emerging from a classical worldview; in fact, classical physics already fails to satisfy this criterion. I've argued elsewhere (Boughn 2019) that measurement prescriptions also lie wholly outside the formalism of classical mechanics.

It is because the descriptions of measurements are wholly removed from the formalism of quantum mechanics that I agree with EPR's conclusion that quantum

---

[12] Howard (1994) has endeavored, successfully I believe, to formalize Bohr's notion of "classical" descriptions of measuring instruments in terms of density matrices and thereby insures consistency with the formalism of quantum mechanics. However, from my pragmatic experimentalist's point of view, I'm satisfied with Bohr's epistemology of experiment. (Boughn 2019)

mechanics "does not provide a complete description of the physical reality".[13] Dyson (2002) made the incompleteness more explicit through four sensible gedanken experiments couched in the ordinary language of measurements and experimental physics that defy quantum mechanical explanation. Quantum mechanics quite simply cannot be applied to all conceivable situations. For Dyson, the dividing line between classical and quantum physics is the same as the dividing line between the past and the future. The wave function constitutes a statistical prediction of future events. After the event occurs, the wave function doesn't collapse rather it becomes irrelevant. The results of past observations, facts, are the domain of classical physics; the probabilities of future events, wave functions, are the domain of quantum physics. Ergo, quantum mechanics does not provide a complete description of nature.

## 3. Final Remarks

I began my essay, as did Carroll, with Feynman's declaration that "nobody really understands quantum mechanics". Carroll refers to it as "Feynman's lament"; however, I seriously doubt that Feynman was bothered by Carroll's quantum riddles. Rather, I suspect that Feynman was simply referring to the wonderful mysteries of quantum mechanics and, perhaps, admonishing those who enter the house of quantum mechanics to leave their notions of classical reality at the door. In fact, immediately following his Messenger Lecture "lament", Feynman (1964) councils us

> So do not take the lecture too seriously, thinking that you really have to understand in terms of some model what I am going to describe, but just relax and enjoy it. I am going to tell you what nature behaves like. If you will simply admit that maybe she does behave like this, you will find her a delightful, entrancing thing.

The seemingly intractable riddles of quantum mechanics weren't the first physicists have encountered. The notion of luminiferous aether confounded physicists beginning in the 18$^{th}$ century. In this case, the "classical" view of reality was that all waves, including light, require physical media in which wave disturbances propagate. After Maxwell, our notion of reality expanded to include dynamic electromagnetic fields

---

[13] On the other hand, I freely admit that I'm not entirely sure what it means for any theory to provide a complete description of physical reality. (see Boughn 2019)

and the notion of an aether medium was discarded.  A more recent example is provided by the nature of time in the context of the special theory of relativity.  In this case, the classical notion of absolute time leads to such unanswerable questions as whether or not two events occur simultaneously.  Ultimately, we had to abandon the notion of absolute time and accept a new reality as prescribed by relativity theory.  This latter conundrum was resolved much more timely, within a few years of Einstein's seminal 1905 paper.  The point is that new phenomena often require new points of view that end up changing our notions of reality.  Embracing these changes leads to a better understanding of the natural world.  So my advice to Carroll is, embrace the wonderful quantum world and the quantum reality it implies.  Do not try to shoehorn the quantum world into a classical reality.

Finally, I absolutely support the pursuit of deep understanding of our physical theories, a tried and true endeavor that has been crucial for our quest of understanding the physical world around us.  The appearance of new phenomena, the unification of theories, and even the pursuit of elegance and simplicity are all legitimate reasons for such an avocation and the successes have been spectacular: spin, superconductivity, neutrinos, anti-matter, black holes, gravitational waves, and many, many more.  The purpose of my essay is simply to point out that attempts to understand quantum mechanics by imposing a classical reality on what are essentially quantum phenomena inevitably leads one down a rabbit hole.  Why not "relax and enjoy it"?

## Acknowledgements

As always I thank my two muses, Freeman Dyson and Marcel Reginatto, for encouraging my pursuit of understanding the foundations of quantum mechanics.

## References

Boughn, S., "On the Philosopical Foundations of Physics: An Experimentalist's Perspective", arXiv:1905.07359 [physics.hist-ph; quant-ph] (2019).

Ballentine, L., "The Statistical Interpretation of Quantum Mechanics", *Rev. Mod. Phys.* **42**, 358-381 (1970).

...

Bell, J., "On the Einstein Podolsky Rosen Paradox", *Physics* **1**, 195–200 (1964).

Bell, J., 1975, "The theory of local beables", presented at the sixth GIFT seminar, appears in *Speakable and Unspeakable in Quantum Mechanics* (Cambridge Univ. Press, Cambridge 2004).

Bohr, N., *Essays 1958/1962 on Atomic Physics and Human Knowledge* (Ox Bow Press, Woodbridge, CT 1963).

Camilleri, K., and Schlosshauer, M., "Niels Bohr as philosopher of experiment: Does decoherence theory challenge Bohr's doctrine of classical concepts?", *Studies in History and Philosophy of Modern Physics* **49**, 73–83 (2015).

Carroll, S., "Even Physicists Don't Understand Quantum Mechanics: Worse, they don't seem to want to understand it.", *NY Times*, Sunday Review, Sept. 7, 2019.

Dyson, F., "Thought experiments—In Honor of John Wheeler", Contribution to Wheeler Symosium, Princeton, 2002, reprinted in *Bird and Frogs* by Dyson, F. pp. 304–321 (World Scientific, Singapore 2015).

Einstein, A., "Physics and Reality", *Journal of the Franklin Institute* **221**, 349-82 (1936).

Einstein, A., Podolsky, B., and Rosen, N., "Can quantum–mechanical description of reality be considered complete?", *Phys. Rev.* **47**, 777–780 (1935).

Feynman, R., *The Character of Physical Law*, The 1964 Messenger Lectures at Cornell University (The Modern Library, New York 1994).

Howard, D., "What Makes a Classical Concept Classical? Toward a Reconstruction of Niels Bohr's Philosophy of Physics", in *Niels Bohr and Contemporary Philosophy*, eds. J. Faye and H. Forse, (Kluwer Acad. Pub. 1994).

Howard, D., "Revisiting the Einstein–Bohr dialogue", *Iyyun: The Jerusalem Philosophical Quarterly* **56**, 57–90 (2007).

Pais, A., *Niels Bohr's Times* (Clarendon Press, Oxford 1991).

Schrödinger, E., "Die gegenwärtige Situation in der Quantenmechanik", *Naturwissenschaften* **23**: 807-812 (1935) [English translation in: J. Trimmer, *Proc. Amer, Phil. Soc.* **124**, 323-38, 1980].

Stapp, H., "The Copenhagen interpretation", *Am. J. Phys.* **40**, 1098–1116 (1972).